\documentclass[twocolumn, showpacs, prl]{revtex4}

\usepackage{graphicx}

\usepackage{dcolumn}

\usepackage{bm}

\begin{document}

\title{Dependence of persistent gaps at Landau level crossings on relative spin}

\date{\today}

\author{K.\ Vakili}

\author{T.\ Gokmen}

\author{O.\ Gunawan}

\author{Y.P.\ Shkolnikov}

\author{E.P.\ De Poortere}

\altaffiliation{Current address: Department of Physics, Columbia
University, New York, NY 10027}

\author{M.\ Shayegan}

\affiliation{Department of Electrical Engineering, Princeton
University, Princeton, NJ 08544}

\begin{abstract}

We report measurements of the quantum Hall state energy gap at
avoided crossings between Landau levels originating from different
conduction band valleys in AlAs quantum wells. These gaps exhibit
an approximately linear dependence on magnetic field over a wide
range of fields and filling factors. More remarkably, we observe
an unexpected dependence of the gap size on the relative spin
orientation of the crossing levels, with parallel-spin crossings
exhibiting larger gaps than antiparallel-spin crossings.

\end{abstract}


\maketitle

In multi-component, two-dimensional (2D) quantum Hall (QH)
systems, the presence of discrete electronic degrees of freedom
can significantly alter the electron-electron interaction.
Recently, there has been increasing interest in the interplay
between different discrete degrees of freedom in these systems.
For example, the spin can influence the correlated $\nu$ = 1 phase
in a double layer system \cite{spielman05}, or change the ground
state symmetry of QH ferromagnetic phases at crossings of Landau
levels (LLs) that originate from different confinement subbands in
a wide GaAs quantum well (QW) \cite{muraki01}. The interplay
between spin and conduction band valley has also been
demonstrated, such as through the dependence of spin
susceptibility on valley degeneracy in AlAs QWs
\cite{shkolnikov04} or the dependence of valley splitting gaps in
SiGe/Si/SiGe QWs on the spins of the valley levels \cite{lai06}.

We studied crossings between the LLs of one valley with the lowest
LL of another valley in 2D electrons in AlAs QWs. This is similar to
the experiments of Ref. \cite{muraki01} where confinement subband
crossings were studied, except we are able to vary the magnetic
field position of the valley crossings, allowing us to study the
field dependence of persistent gaps at those crossings. Also, the
transfer of charge between different confinement subbands has a
capacitive energy component due to charge redistribution along the
confinement direction, whereas the valley degree of freedom has no
spatial charge redistribution and, hence, no capacitive energy
associated with it, yielding a somewhat simpler system. We observe
an approximately linear dependence of the persistent gaps at
crossings of different valley LLs on magnetic field and,
surprisingly, a dependence on the relative spin of the levels, with
larger gaps for the parallel spin case than for antiparallel.

Bulk AlAs has an indirect band gap with conduction band minima at
the six equivalent X-points of the Brillouin zone. The Fermi
surface of electrons, therefore, consists of three (six half)
prolate ellipsoids (or valleys), each with its major axis oriented
parallel to the crystal axis along which the valley center is
displaced. Electrons occupying these anisotropic valleys have a
longitudinal band effective mass of {\it m}$_{l}$ = 1.04 and
transverse masses of {\it m}$_{t}$ = 0.21, in units of the vacuum
electron mass. When confined to AlAs QWs thicker than 55 $\AA$, as
is the case for our samples, the strain associated with lattice
mismatch between the AlAs QW and the GaAs substrate on which it is
grown causes only the valleys oriented in the plane of the QW to
be occupied up to the highest accessible densities. We will refer
to these in-plane valleys as X and Y.  The band effective mass,
associated with the density of states and the cyclotron energy,
for these valleys is given by {\it m}$_b$ = ({\it m}$_l${\it
m}$_t$)$^{1/2}$ = 0.47.

We measured two samples (A and B) from a wafer containing a 110
$\AA$ AlAs QW \cite{depoortere02}.  Each sample was patterned in a
Hall bar geometry and fitted with metallic front and back gates to
vary the charge density, {\it n}, from 1.5 to 8.2 x 10$^{11}$
cm$^{-2}$. Typical mobilities are near 26 m$^2$/Vs. Ohmic contacts
were made by depositing AuGeNi and alloying in a reducing
environment. The samples were thinned to approximately 150 $\mu$m
and glued to a piezoelectric actuator to allow for {\it in situ}
application of a symmetry-breaking, in-plane strain that varies
the energy splitting of the two occupied valleys, as described
previously \cite{shayegan03}.  We employed standard low-current,
low-frequency lock-in techniques, and the samples were cooled in a
pumped $^3$He refrigerator with a base temperature, {\it T}, of
0.3 K that was equipped with a single-axis tilting stage, allowing
for variation of the angle, $\theta$, between the sample normal
and the applied magnetic field, {\it B}. We refer to the component
of {\it B} perpendicular to the QW plane as {\it B}$_{\bot}$.

\begin{figure}
\centering
\includegraphics[scale=0.7]{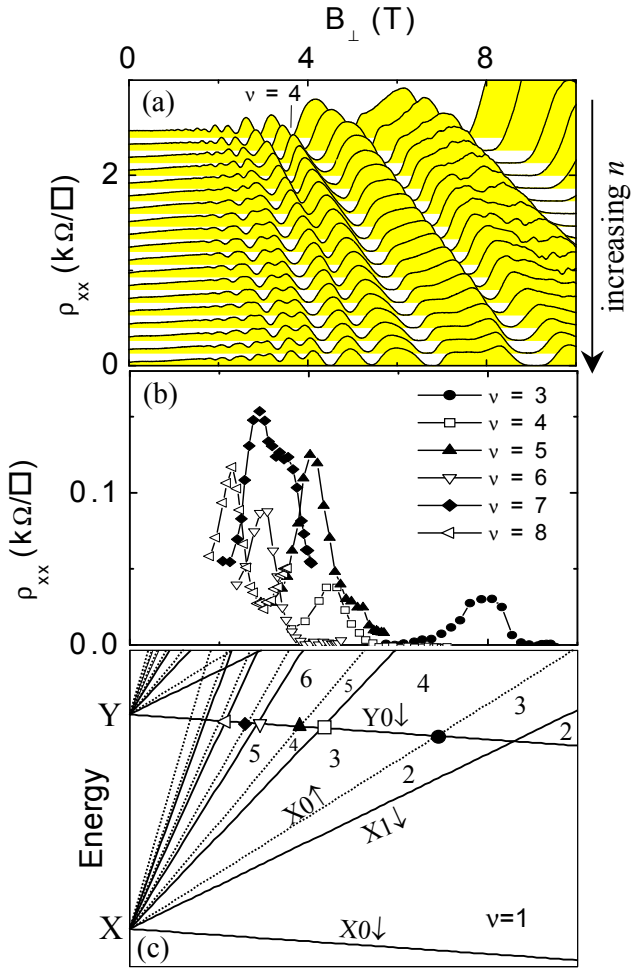}
\caption{(a) $\rho_{xx}$ vs. {\it B}$_{\bot}$ at {\it T} = 0.3 K
as {\it n} increases from 3.5 (top) to 6.6 x 10$^{11}$ cm$^{-2}$
(bottom). Traces are vertically offset for clarity. (b)
$\rho_{xx}$ at integer $\nu$ extracted from the traces in (a). (c)
Simulated energy level diagram (energy units are arbitrary) for
the data in (a) and (b), showing reasonable agreement between the
indicated LL crossings and the extrema in $\rho_{xx}$ at integer
$\nu$. Solid (dotted) lines are majority (minority) spin LLs, and
parallel and antiparallel LL crossings are shown by open and
closed symbols respectively. Our notation indicates the energy
levels' valley, LL index, and spin.}
\end{figure}

To locate and study the crossings of the LLs corresponding to the
two occupied valleys, we employed two different methods. The first
is demonstrated in Fig. 1 for sample A.  We set the valley
splitting to a particular, fixed value by applying a fixed voltage
to the piezo \cite{fixed}. The energy level (fan) diagram for the
system then looks qualitatively like that shown in Fig. 1(c), with
the LLs of the majority (X) valley crossing the lowest LL of the
minority (Y) valley at various {\it B}$_{\bot}$. Some of the
filling factors, $\nu$, defined as the number of occupied LLs
below the Fermi energy, {\it E}$_F$, are indicated in the diagram
as well. Since the magnetic field corresponding to a given $\nu$,
{\it B}$^{\nu}_{\bot}$ = {\it hn}/{\it e}$\nu$, depends on {\it
n}, we can change the energy gap for an integer QH state at a
particular $\nu$ by changing {\it n}. This gap is a local minimum
when {\it B}$^{\nu}_{\bot}$ falls at the LL crossing for that
$\nu$. Consequently, $\rho_{xx}$ at $\nu$ should be a maximum for
a particular value of {\it n}, thereby revealing the field
position of the LL crossing. In Fig. 1(a), we show traces of
$\rho_{xx}$ vs. {\it B}$_{\bot}$ as {\it n} is increased from 3.5
(top) to 6.6 x 10$^{11}$ cm$^{-2}$ (bottom). Tracking $\rho_{xx}$
at different integer $\nu$, we obtain the curves shown in Fig.
1(b), with the peak of each curve indicating the position of the
LL crossing for that $\nu$.  It can be shown that the relevant
quantities that determine the positions of these crossings are the
spin and valley susceptibilities of each valley and the applied
piezo voltage (the overall energy scale is arbitrary).  To
generate the diagram in Fig. 1(c), we  employ recent measurements
of the valley susceptibility in these samples \cite{fixed}, and
use equal spin susceptibilities for the two valleys, with its
value a fitting parameter \cite{spin}. The spin susceptibility is
proportional to {\it g}*{\it m}*, where {\it g}* and {\it m}* are
the (interaction enhanced) Land\'{e} g-factor and effective mass,
and we have used {\it g}*{\it m}* = 2.5. This is enhanced by a
factor of 2.7 above {\it g}$_b${\it m}$_b$ ({\it g}$_b$ = 2) and
is in reasonable agreement with measured values in the two-valley
limit \cite{shkolnikov04}. The resulting fan diagram successfully
reproduces the observed crossings \cite{nu3}.

The second method that we have employed to locate crossings
between the X and Y LLs involves fixing {\it n} and {\it
B}$_{\bot}$ so that $\nu$ is fixed at an integer value.
$\rho_{xx}$ is then monitored as the piezo voltage is swept such
that the 2D electron system is taken from the strongly
single-valley limit towards the two-valley limit. Along the way,
the desired crossing is revealed as a local maximum in
$\rho_{xx}$.  For both methods, we have confirmed that the
$\rho_{xx}$ maxima correspond to minima of the energy gaps by
measuring these gaps at and away from the LL crossings.

\begin{figure}
\centering
\includegraphics[scale=0.39]{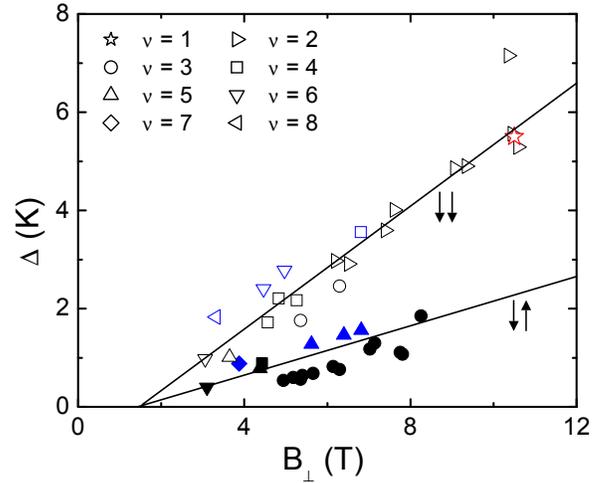}
\caption{Energy gaps ($\Delta$) at LL crossings as derived from
temperature-dependent activation measurements.  Different symbols
correspond to different $\nu$ as indicated, and open (closed)
symbols are parallel (antiparallel) spin crossings. Black points
are for sample A, blue for sample B, and red for a point derived
from the data of Ref. \cite{shkolnikov05}. The lines are guides to
the eye.}
\end{figure}

Our main finding is first hinted at in Fig. 1(b). The maximum
value of $\rho_{xx}$ at integer $\nu$, attained at the LL
crossings, {\it oscillates} with $\nu$, with odd $\nu$ exhibiting
larger $\rho_{xx}$ peaks than even $\nu$. From our fan diagram in
Fig. 1(c), we see that the LL crossings at even $\nu$ occur
between LLs with parallel spin while those at odd $\nu$ have
antiparallel spin. We have further investigated this
quantitatively by performing temperature dependence measurements
of the persistent, activated gaps ($\Delta$) at the LL crossings
with {\it T} typically varying between 0.3 and 6 K. The crossings
between LLs with the parallel spin occurring at even $\nu$ indeed
appear to fall on a different, higher $\Delta$ branch than those
between antiparallel spin LLs.

\begin{figure}
\centering
\includegraphics[scale=0.43]{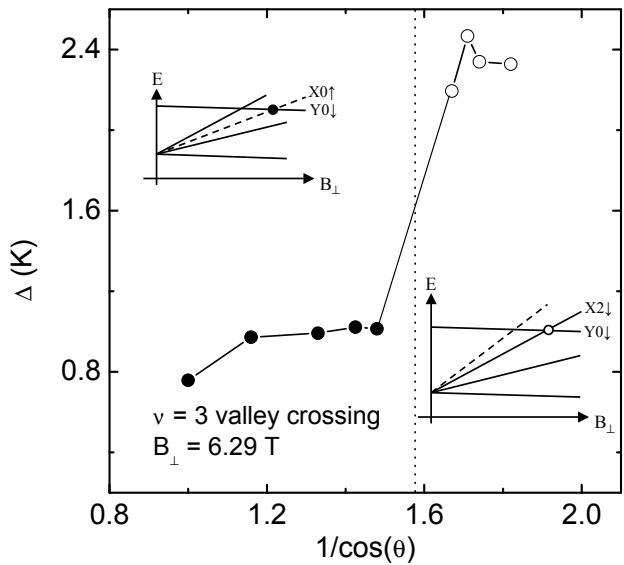}
\caption{Energy gap, $\Delta$, at the $\nu$ = 3 LL crossing as
sample A is tilted through the first coincidence angle at fixed
{\it n} and {\it B}$_{\bot}$. The vertical dotted line indicates
the coincidence angle, and schematic fan diagrams on either side
of this coincidence are shown as insets.}
\end{figure}

To further test the dependence of $\Delta$ on the relative spins
of the crossing LLs, we tilted the sample with respect to {\it B}.
Since the cyclotron energy, {\it E}$_C$ $\propto$ {\it
B}$_{\bot}$, depends on the perpendicular component of the field
while the Zeeman energy , {\it E}$_Z$ $\propto$ {\it B}, depends
on the {\it total} field, tilting the sample causes majority
(down) spin LLs to drop in energy while minority (up) spin levels
rise for a given value of {\it B}$_{\bot}$.  The result is that
the different spin LLs can be brought to and driven past energetic
coincidence \cite{fang68}, thus changing the relative spin of the
LLs at the crossings. As an example, we show the gap at the $\nu$
= 3 crossing as the sample is tilted in Fig. 3. Initially, the
crossing at $\nu$ = 3 occurs between opposite spin LLs
(Y0$\downarrow$ and X0$\uparrow$) but, as the sample is tilted,
the relative spin of the crossing LLs at $\nu$ = 3 changes from
antiparallel to parallel (Y0$\downarrow$ and X2$\downarrow$) when
the first coincidence angle at 1/cos($\theta$) $\simeq$ 1.6 is
passed. It is apparent from this data that the dependence of the
size of $\Delta$ on the relative spins of the crossing LLs is
preserved with tilt. The values of $\Delta$ for sample A in tilted
field are included in Fig. 2.

One factor that we have neglected so far is the relative LL
indices of the crossing LLs.  If the finite gaps that we observe
at LL crossings result from exchange-interaction induced QH
ferromagnetic phases, then the symmetry of the ferromagnetic
ground states and excitations, and hence the size of $\Delta$,
would depend in general on the relative LL indices of the crossing
levels \cite{jungwirth00}. When these indices are different, the
difference in wavefunction shape yields an easy-axis symmetry
while, for same indices, the symmetry is isotropic as determined
by the spin and/or valley degrees of freedom. This has been
confirmed by recent observations of valley skyrmions at $\nu$ = 1
in valley degenerate AlAs QWs \cite{shkolnikov05}. Since the LL
indices can affect the ground state symmetry and excitations of a
QH ferromagnetic state, we must examine whether they are
responsible for determining the sizes of the gaps. From Fig. 1(c),
it is evident that all of the crossings that we have studied in
sample A at $\theta$ = 0 occur between LLs with different LL index
except at $\nu$ = 3, where both levels have LL index {\it N} = 0.
Though there are gaps on the lower branch in Fig. 2 that
correspond to crossings between LLs having different LL indices,
most of the data on this branch come from $\nu$ = 3. Also, when
the sample is tilted, the relative LL indices for the $\nu$ = 3
crossing go from being the same to being different, which could
explain the change in $\Delta$ as the difference in energy between
skyrmion and single-flip excitations. Similarly, the relative LL
indices at the $\nu$ = 4 crossing go from being different to being
the same with tilt, so the reduction in $\Delta$ in that case
could be explained along similar lines.

To distinguish between the effect of LL index and spin on
$\Delta$, we measured the same persistent gaps in a second sample
(B).  The valley splitting in sample B is larger than in sample A
for a given piezo voltage due to residual strain associated with
the cooldown procedure, so that LL crossings at particular $\nu$
occur at higher {\it n} and {\it B}$_{\bot}$ \cite{angle}. This
allows us, for example, to compare $\Delta$ at $\nu$ = 2
(Y0$\downarrow$ and X1$\downarrow$ crossing) with $\nu$ = 5
(Y0$\downarrow$ and X1$\uparrow$ crossing). If we can ignore the
influence of the LLs below and above {\it E}$_F$ for the moment,
then only the relative spin of the crossing levels is different
for $\nu$ = 2 and $\nu$ = 5. The same is true for $\nu$ = 1 and 3
(Y0$\downarrow$ with X0$\downarrow$ or X0$\uparrow$ respectively)
and for $\nu$ = 4 and 7 (Y0$\downarrow$ with X2$\downarrow$ or
X2$\uparrow$ respectively). We have included the measured values
of $\Delta$ for sample B in Fig. 2 as blue symbols, as well as one
point (red star) corresponding to the persistent gap at $\nu$ = 1
derived from the data in Ref. \cite{shkolnikov05}. These
additional data points confirm that it is indeed the relative {\it
spin} of the crossing LLs that determines which of the two
branches in Fig. 2 $\Delta$ falls on.

The reason for the simple yet unexpected dependence of $\Delta$ on
the relative spins of the crossing LLs is not entirely clear.  In
a previous study of persistent gaps at crossings of different
confinement subband LLs in a wide GaAs QW, a dependence on the
relative spin of crossing LLs was observed and explained as a
consequence of the formation of different QH ferromagnetic states
\cite{muraki01}. That explanation relied on the difference in
charge distribution along the confinement direction for the two
confinement subband states, whereas such a factor does not apply
for the X and Y valley subbands in our experiments.  Furthermore,
in our measurement the behavior of $\Delta$ away from the LL
crossings is qualitatively similar for parallel and antiparallel
spin crossings, both exhibiting a smooth variation reminiscent of
an anticrossing.

We propose two possible explanations for the dependence of
$\Delta$ on relative spin that is observed in Fig.\ 2.  The first
explanation assumes that the persistent gaps are, in fact, a
consequence of the formation of QH ferromagnetic states. In this
case, $\Delta$ corresponds to the exchange energy cost associated
with the creation of many-body excitations and, thus, depends on
the strength of interaction between the electrons. The fact that
$\Delta$ increases monotonically with {\it B}$_{\bot}$ is
consistent with this interpretation, and the same idea underlies
the explanation for {\it B}$_{\bot}$ dependent valley splitting in
nominally valley degenerate 2DESs that has been reported
previously \cite{shkolnikov02}.  The variation of $\Delta$ with
relative spin, then, may be a consequence of the difference in the
strength of screening of the Coulomb interaction.  Lowest-order
(Thomas-Fermi) screening depends only on the charges at {\it
E}$_F$, however there exist higher-order screening processes
associated with the promotion of charges from LLs below {\it
E}$_F$ to unoccupied LLs above.  When the X valley LL at the
crossing at {\it E}$_F$ has minority spin, then invariably there
exists a majority spin LL with the same LL index below {\it
E}$_F$, and the electrons in this level could screen the Coulomb
interaction in the corresponding minority spin LL effectively
thanks to the same wavefunction shape. On the other hand, when the
X valley LL at the crossing has a majority spin, there is no other
LL below {\it E}$_{F}$ with the same LL index, and thus the
screening of the Coulomb potential by the charges in these
low-lying LLs is less effective.  This could produce the observed
effect.

A second possibility is that some part of the finite gap results
from single-particle mixing between the coincident LLs.  Such
mixing, associated with the existence of finite off-diagonal terms
in the two-component Hamiltonian, yields an anti-crossing with a
finite gap determined by the magnitude of the off-diagonal terms.
Though mixing between X and Y valley states in AlAs is possible
\cite{fu93}, it would be suppressed when the relative spins of the
coincident LLs are different, since single-particle mixing between
different spin states can only be facilitated by scattering from
such entities as magnetic impurities or nuclear spins.  Thus, in
this interpretation, the difference in energy between the upper
and lower branches in Fig. 2 would correspond to the mixing
strength of the X and Y valley states.  There are, however,
several puzzling aspects in this interpretation.  First, there is
no obvious reason why the valley mixing should depend on {\it
B}$_{\bot}$, as the behavior of $\Delta$ in Fig. 2 suggests it
would.  Second, there are still finite gaps for crossings of
different spin LLs, suggesting that either there is some mixing
between opposite spin states, or that there is an additional
contribution to the gaps from other effects, such as the formation
of QH ferromagnetic states discussed above. Finally, there is a
question as to whether single-particle anticrossings are even
possible in the QH regime.  That would require the single-particle
mixing strength of the two components at each guiding center to
not vary significantly over all guiding centers and, since the
mixing is almost certainly disorder-dependent, this is unlikely.

In summary, we have observed finite gaps at crossings of LLs
originating from two different conduction band valleys.  The size
of these gaps appears to depend linearly on {\it B}$_{\bot}$ and
also exhibits a dependence on the relative spin of the crossing
levels, with parallel-spin crossings exhibiting larger gaps than
antiparallel-spin crossings.  The dependence on relative spin may
be a consequence of variation in the strength of screening of the
Coulomb interaction for these two cases, or it may signal the
existence of single-particle mixing between the different valley
states.

We thank the NSF for financial support, and A.H.\ MacDonald for
illuminating discussions.

\break


\begin{references}
\bibitem{spin}
The assumption of equal susceptibilities for the two valleys means
that only the positions of crossings between LLs of opposite spin
are affected by our choice of our fitting parameter, the spin
susceptibility.
\end{references}
\end{document}